\RequirePackage{fix-cm}
\documentclass[smallextended,natbib]{svjour3}   
\smartqed  
\usepackage{amsmath}
\usepackage{amssymb}
\usepackage{graphicx}
\usepackage{wrapfig}
\usepackage{subcaption}
\usepackage{makecell}
\usepackage{booktabs}
\usepackage{multirow}
\usepackage{xspace}
\usepackage{xcolor}
\usepackage{colortbl}
\usepackage{enumitem}
\usepackage{listings}
\usepackage{tcolorbox}
\usepackage{scalerel}
\usepackage{tikz}
\usepackage{hyperref}
\hypersetup{
    colorlinks=true,
    linkcolor=blue,    
    citecolor=blue,    
    urlcolor=blue,     
    linktocpage=true
}

\definecolor{gold}{RGB}{255,215,0}
\definecolor{silver}{RGB}{192,192,192}
\definecolor{orange}{RGB}{255,165,0}

\renewcommand{\cite}{\citep}

\newcommand\pick[1]{{\textcolor{black}{#1}}}

\definecolor{mygreen}{rgb}{0.0, 0.5, 0.0}

\newcommand\hieu[1]{{\textcolor{black}{#1}}}

\newcommand{\tool}{\textsc{XAgent}\xspace}

\newcommand{\claudelogo}{\scalebox{1}{\scalerel*{\includegraphics{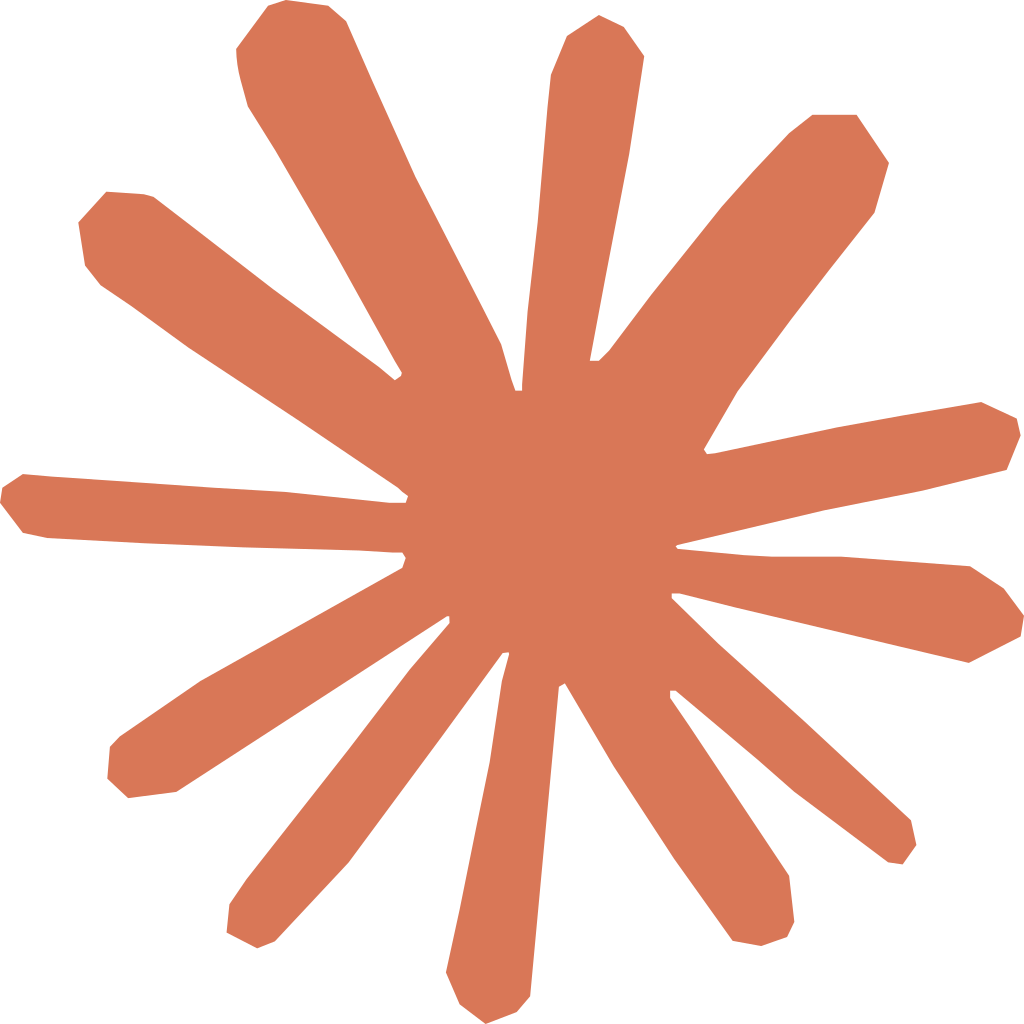}}{\textrm{C}}}\xspace}

\definecolor{fillblue}{HTML}{b9d7f1}
\definecolor{borderblue}{HTML}{2e3192}

\newcommand{\circled}[1]{%
  \tikz[baseline=(char.base)]{
    \node[
      circle,
      draw=borderblue,
      fill=fillblue,
      inner sep=0.5pt,
      minimum size=1.1em,
      line width=0.4pt,
      text height=1.2ex,
      text depth=.25ex
    ] (char) {\scriptsize #1};
  }%
}

\newcommand{\finding}[2]{
    \begin{tcolorbox}[
        colback=gray!5,
        colframe=gray!60!black,
        boxrule=1pt,
        arc=1mm,
        left=1pt, right=1pt,
        top=1pt, bottom=1pt,
    ]
        \textbf{#1} #2
    \end{tcolorbox}
}

\newcommand{\smallsection}[1]{\noindent\underline{\noindent {\bf #1}.}\hspace{1mm}}

\journalname{Empirical Software Engineering}
\makeatletter
\def\makeheadbox{}
\makeatother

\begin{document}

\title{\tool: eXecution-guided Agentic AI for Effective Localization and Resolution of GitHub Issues}
\titlerunning{\tool: eXecution-guided Agentic AI for GitHub Issues}

\author{Hieu Huynh \and
        Patanamon~Thongtanunam \and
        Michael Fu \and
        Bach Le \and
        Kla~Tantithamthavorn
}
\authorrunning{Huynh et al.}

\institute{Hieu Huynh \at
              School of Computing and Information Systems, The University of Melbourne, Melbourne, Australia \\
              \email{minhhieuh@student.unimelb.edu.au}
           \and
           Patanamon Thongtanunam \at
              School of Computing and Information Systems, The University of Melbourne, Melbourne, Australia \\
              \email{patanamon.t@unimelb.edu.au}
           \and
           Michael Fu \at
              School of Computing and Information Systems, The University of Melbourne, Melbourne, Australia \\
              \email{michael.fu@unimelb.edu.au}
           \and
           Bach Le \at
              School of Computing and Information Systems, The University of Melbourne, Melbourne, Australia \\
              \email{bach.le@unimelb.edu.au}
           \and
           Kla Tantithamthavorn \at
              Monash University, Clayton, Australia \\
              \email{chakkrit@monash.edu}
}

\date{Received: date / Accepted: date}

\maketitle

\begin{abstract}
Agentic AI has enabled capabilities in leveraging Large Language Models (LLMs) to autonomously resolve repository-level GitHub issues. 
However, due to the reliance on limited static description of issues, existing agentic approaches suffer from incorrect localization and incomplete validation.
Solely relying on this information can bias LLM reasoning toward the narrow scope of the issue description, leading to incomplete patches that fail to address the underlying issue. 
In this paper, we present \tool, an execution-guided agentic framework that analyzes dynamic
behavior and additional program context to localize and validate issues. 
The experimental results on the SWE-bench-lite dataset demonstrate that \tool outperforms other existing approaches, achieving a resolve rate of 62.0\% and a function localization accuracy of 72.8\%, while maintaining cost efficiency. Our analysis further shows that \tool successfully resolves 7 additional issues that the top existing baselines fail to address.
This work highlights a shift from static, description-oriented patch generation toward dynamic execution-guided issue resolution, opening new opportunities for LLM-based coding agents to achieve more robust and generalizable software maintenance.

\end{abstract}

\section{Introduction}
\label{section:intro}

Agentic AI has demonstrated potential to address various software engineering tasks including resolving real-world GitHub issues (e.g., AutoCodeRover \cite{zhang2024autocoderover}, Agentless \cite{agentless}, ExpeRepair~\cite{experepair}, SWE-agent \cite{yang2024sweagent}, and CoSil~\cite{jiang2025cosil}).
Such agentic approaches leverage Large Language Models (LLMs) as agents to autonomously perceive the input environments (e.g., GitHub repositories), reason (e.g., generating a coding plan), and autonomously perform actions (e.g., open files, run bash scripts) to achieve an ultimate goal (e.g., resolve a GitHub issue at a repository-level).
Typically, such approaches follow a three‑stage workflow: (1) reproduction, which recreates the reported buggy behavior by generating test cases that verify the described issue; (2) localization, which identifies the relevant files, functions, and code regions responsible for the buggy behavior; (3) patch generation \& validation, which produces candidate fixes for localized components and evaluates whether the proposed patch resolves the issue through execution of the reproduction and regression test.




However, existing agentic approaches are still ineffective in resolving GitHub issues.
For example, ExpeRepair \cite{experepair}, which is \#1 in the SWE-Bench-lite benchmark dataset can correctly localize 70.7\% of the buggy functions and correctly resolve only 60.3\% of the GitHub issues, which is still far from perfect.
We suspect that the ineffectiveness has to do with the over-reliance on static issue descriptions for the localization and the test validation stages in the existing agentic approaches.
In particular, we identified the following two major limitations.

\begin{itemize}[leftmargin=*]
    \item \textbf{Limitation~1: Incorrect Localization.} The localization stage of the existing approaches \cite{yang2024sweagent,wang2024openhands,experepair,agentless} relies on the semantic similarity of keywords that appear in the issue descriptions and the files and functions within the whole repository (i.e., keyword search). 
    However, as issue descriptions are generally noisy or incomplete, the sole-reliance on static issue descriptions may bias the LLMs toward functions that are irrelevant or incorrect to the given issue description, leading to an incorrect localization by the~existing~approaches.
    
    \item \textbf{Limitation 2: Incomplete Validation.} Similarly, the test validation stage of the existing approaches \cite{agentless,experepair} solely uses the issue description to generate a set of validation tests without considering the broader program context. As a result, the validation process often overfits the narrow scope defined by the issue description, causing the generated patches to pass the validation test but fail to address the underlying issue. These limitations make existing approaches prone to producing patches that overfit to the described issue (i.e., overfitting patches~\cite{10.1145/2786805.2786825, 10.1145/3180155.3182536}), ultimately restricting their effectiveness in real-world scenarios.
\end{itemize}

In this paper, we present \tool, an execution-guided agentic framework for GitHub issue resolution that analyzes dynamic behavior during program execution to localize functions and leverages program context to generate more reliable validation tests.
Inspired by differential analysis \cite{zeller2009programs}, \tool{} identifies the buggy functions by comparing failing and passing executions of the program.
Specifically, it performs localization by identifying suspicious functions that appear only in failing execution but not in the passing one.
To prevent the validation test from overfitting to the issue description, we develop a context-aware test augmentor that leverages additional program context (e.g., relevant files and functions) to broaden the scope of validation beyond what is described in the issue.
This enables \tool{} to  generate edge-case and related-component tests, ensuring that the generated patches address the underlying issue.


Experimental results show that \tool achieves state-of-the-art performance on the SWE-bench-lite dataset, surpassing top baselines with a resolve rate of 62.0\%. \tool also demonstrates highly effective bug localization, achieving SOTA function localization accuracy of 72.8\%,  while maintaining high efficiency at a cost of approximately \$1.56 per issue, which is 37\% cheaper than the prior SOTA approach~\cite{experepair}. Additionally, our analysis reveals that \tool uniquely resolves 7 complex instances that other top approaches (i.e., ExpeRepair \cite{experepair}, Refact Agent \cite{refact}, SWE-Agent \cite{yang2024sweagent}) failed to fix.
The results show that \tool can accurately localize and cost-effectively generate correct patches to resolve the underlying issues.
This work highlights a shift from static, description-oriented patch generation toward dynamic, execution-guided issue resolution, opening new opportunities for LLM-based agents to achieve more robust and generalizable software~maintenance.

\textbf{Novelty \& Contributions.} To the best of our knowledge, we are the first to propose:
\begin{itemize}[leftmargin=*]

    \item \textbf{\tool:} A novel execution-guided agentic framework for GitHub issue resolution that shifts from static, description-oriented approaches to a dynamic, execution-guided and context-aware approach.
    
    
    \item A novel \textbf{execution-guided localization} method using differential analysis (buggy vs. non-buggy) and tree edit distance to localize bugs by analyzing execution divergence, effectively addressing the limitation of existing static bug localization. 
    
    \item A \textbf{context-aware test augmentor} that generates broader validation tests targeting edge cases, effectively extending validation coverage beyond the issue description.
    
    \item A \textbf{comprehensive empirical evaluation} on SWE-Bench-Lite shows that \tool sets a new state of the art, achieving the highest Resolve Rate (62.0\%) and function-level localization accuracy (72.8\%) while reducing cost by 37\%. Furthermore, \tool successfully resolves 7 issues that top existing baselines (e.g., SWE-Agent, EXPEREPAIR, Refact) failed to address.

    \item We conduct \textbf{an ablation study} to verify our key design for \tool, demonstrating that the execution-guided localization and context-aware test augmentor together contribute up to 9\% relative improvement in the overall performance.
\end{itemize}

\section{Background \& Related Work}
\label{section:related_works}


\textbf{Background.} Software issue resolution at the repository level is a core software maintenance activity that requires understanding, localizing, and fixing real-world bugs in complex codebases based on issue descriptions provided by users. Given an issue report, developers typically reproduce the problem to confirm the issue, localizing the bug, implement a patch, and validate the fix through testing. 
Recent advances in LLMs have significantly transformed this process by serving as powerful reasoning and code generation components in software engineering, enabling a shift from manual workflows toward increasingly autonomous solutions. To further enhance effectiveness, agentic AI has been adopted to enable LLMs to interact directly with code repositories through tool use, allowing them to iteratively plan, execute, and adapt their strategies based on environmental feedback.
Existing agentic methods for repository-level issue resolution typically organize the workflow into three main stages: (1) reproduction, (2) localization, and (3) patch generation and validation.
Below, we outline how LLMs are used at each stage of software issue resolution within a code repository.

\textbf{Reproduction} is the stage for reproducing the behavior as users described in the issue description.
Typically, existing approaches \cite{jiang2025cosil,yang2024sweagent,experepair,agentless,yu2025orcaloca,wang2024aegis,nashid2025issue2test,libro} leverage an LLM to generate the reproduction test based on the issue description and the code context of the repository.
This reproduction test then executes the program to verify the existence of the reported issue.

\textbf{Localization} aims to identify the specific files and functions to be fixed \cite{jiang2025cosil,chakraborty2025blaze}. 
Existing work \cite{yang2024sweagent,wang2024openhands,experepair,agentless} relies on LLMs to extract keywords from issue descriptions and search for files and functions to be fixed within the whole repository.
For example, AGENTLESS \cite{agentless} adopts a hierarchical localization strategy by extracting keywords from the issue description, identifying the top-K relevant files, then narrowing down to candidate classes, functions, and finally specific edit locations.
AutoCodeRover \cite{zhang2024autocoderover} also performs localization by keyword search, then further refine it by leveraging the spectrum-based fault localization technique if test suites are available in the repository.
LocAgent \cite{chen2025locagent} localizes bugs by using keywords from the issue description to search through the repo-level code graph.
OrcaLoca \cite{yu2025orcaloca} analyzes the error trace provided in the issue description to identify the relevant files and functions.
PatchPilot \cite{patchpilot} enhances localization by leveraging the execution information from the reproduction test.
Despite the promising results, these existing works heavily relied on the limited information in the issue description and the reproduction test.
\finding{Limitation 1:}{The limited information in the issue description and the reproduction test may bias the LLMs toward the failure scenario, leading them to misidentify the buggy function.}

\textbf{Patch Generation \& Validation} is the stage when an LLM produces patch candidates for the localized functions and validates whether the patch resolves the issue. 
Typically, existing approaches \cite{yang2024sweagent,wang2024openhands,experepair,patchpilot,yu2025orcaloca} iteratively generate the patches for the localized functions and validate them against the validation tests. These validation tests include the reproduction test to verify that the reported issue has been resolved, as well as regression tests to ensure that the fix does not adversely affect existing functionality.
To effectively achieve this task, many approaches leverage agentic method ~\cite{yang2024sweagent,wang2024openhands,zhang2024autocoderover,experepair} which enable an LLM autonomously to select actions and adjusts its plan.
For example, SWE-agent~\cite{yang2024sweagent} and OpenHands~\cite{wang2024openhands} equip LLMs with tools to interact directly with the coding environment, such as 
viewing files, searching code, and editing source files, enabling them to plan and decide actions based on feedback from the validation test results.
EXPEREPAIR \cite{experepair} further improves the method by integrating a Dual-Memory System that allows LLMs to retrieve knowledge accumulated from previous runs.
In addition, they expand validation tests by generating additional tests from original reproduction tests using different hyper-parameters. 
Despite the effectiveness of the agentic method in generating patches, the validation process is solely based on the issue description without considering the broader program context.

\finding{Limitation 2:}{The validation process often overfits narrow scopes defined by the issue description, causing the generated patches to pass the validation test but fail to address the underlying issue.}

\section{Motivating Examples}
\label{section:motivating}

\begin{figure}[ht]
    \centering
    \begin{minipage}{\linewidth}
        \centering
        
        \begin{subfigure}[b]{\linewidth}
            \centering
            \includegraphics[width=\linewidth]{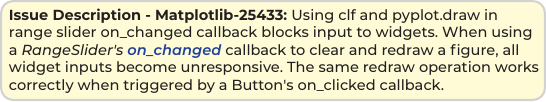}
            \label{fig:issue1}
        \end{subfigure}
        \vspace{-20pt}
        
        \begin{subfigure}[t]{0.49\linewidth}
            \centering
            \includegraphics[width=\linewidth]{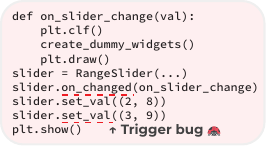}
            \vspace{-16pt}
            \caption{Reproduction Script (Buggy)}
            \label{fig:buggy_tc1}
        \end{subfigure}
        \hfill
        \begin{subfigure}[t]{0.49\linewidth}
            \centering
            \includegraphics[width=\linewidth]{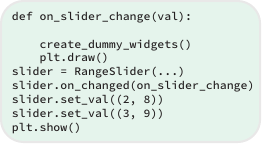}
            \vspace{-16pt}
            \caption{Reference Script (Non-buggy)}
            \label{fig:non-buggy-tc1}
        \end{subfigure}
        
        \begin{subfigure}[t]{0.49\linewidth}
            \centering
            \includegraphics[width=\linewidth, trim=0 0 0 0.38cm, clip]{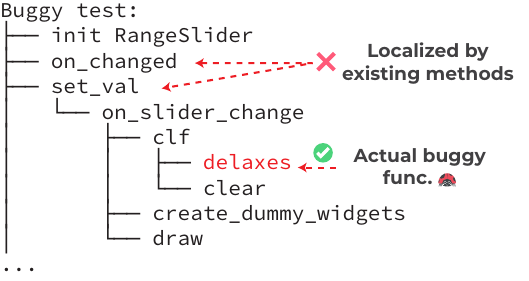}
            \vspace{-16pt}
            \caption{Buggy Execution Trace. \\Existing methods \cite{experepair,refact,yang2024sweagent,yu2025orcaloca}}
            \label{fig:buggy_trace}
        \end{subfigure}
        \hfill
        \begin{subfigure}[t]{0.49\linewidth}
            \centering
            \includegraphics[width=\linewidth, trim=0 0 0 0.38cm, clip]{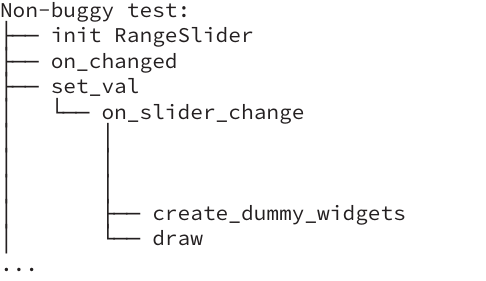}

            \caption{Non-buggy Execution Trace}
            \label{fig:non-buggy_trace}
        \end{subfigure}
        
    \end{minipage}
    \caption{\textit{Matplotlib-25433} Issue.}
    \label{fig:ex1}
\end{figure}
\subsection{Example 1} 
\label{sec:example1}
Fig \ref{fig:ex1} illustrates the issue \textit{Matplotlib-25433} in the SWE-bench-lite dataset. This issue occurs when a user creates a plot with a slider that controls a dynamic value in the graph. The slider is associated with a callback function (i.e., an \texttt{on\_slider\_change} handler) that is triggered whenever the user drags the slider. The problem occurs when this callback calls \texttt{clf()} to clear and redraw the figure: the \texttt{clf()}) function removes all plot elements, including the slider currently being used, without releasing its mouse event. As a result, Matplotlib continues to wait for input from a slider that no longer exists, causing the application to freeze and ignore further mouse interactions. The issue was fixed by modifying the \texttt{delaxes} function (called by \texttt{clf()}) to release all active mouse events before clearing the figure, ensuring that interactive plots remain responsive.


\textbf{Existing approaches.} Refact \cite{refact} modifies the \texttt{on\_changed} as their LLM reasoning tends to follow the keyword \texttt{on\_changed} mentioned in the issue description. On the other hand, other approaches \cite{experepair,yang2024sweagent,yu2025orcaloca} localize the \texttt{set\_val} function as this function triggers the error in the reproduction script. However, the actual bug resides~in~the \texttt{delaxes} function, which lies deeper in the runtime execution traces and is not explicitly mentioned in the issue description.

Indeed, the functions can be efficiently localized based on a differential strategy~\cite{zeller2009programs}: \textit{"code that is executed only in failing runs is more likely to contain the defect than code that is always executed"}. This inspires us to focus on contrasting failing behavior against a passing one.
In other words, rather than solely focusing on the buggy scenario (Fig \ref{fig:buggy_tc1}), a non-buggy scenario (Fig \ref{fig:non-buggy-tc1}) that demonstrates the correct and expected behavior could also provide important information for bug localization.
Specifically, when we compare the execution traces between these two scenarios, specific functions that are invoked exclusively in the buggy scenario will be highlighted.
Hence, these functions are potentially buggy.

For example, Fig ~\ref{fig:buggy_trace}-\ref{fig:non-buggy_trace} presents simplified buggy and non-buggy traces. 
We can clearly see that \texttt{set\_val} and \texttt{on\_changed} functions appear in both traces, only \texttt{clf}, \texttt{delaxes}, and \texttt{clear} functions appears uniquely in the buggy trace.
These functions precisely match the actual buggy function identified by the developer.
Moreover, this comparison between the buggy and non-buggy execution also helps reduce the search space of the functions.
Specifically, there are 756 functions in the buggy trace and 758 functions in the non-buggy trace. Comparing the two traces reveals only \textbf{15} functions that are uniquely invoked in the buggy trace.
This supports our intuition that \textbf{the behavioral differences between the buggy and non-buggy scenarios might reduce the search space significantly and could reveal the functions where the issue is likely located.}

\finding{Key Idea 1: Execution-guided Localization.}{Comparing buggy and non-buggy executions could reveal functions with divergent behaviors, which are likely potentially buggy~functions.}

\subsection{Example 2} 
\begin{figure}[t]
    \centering
    \begin{minipage}{\linewidth}
        \centering

        \begin{minipage}[b]{0.5\linewidth}
            \centering
            
            \begin{subfigure}[b]{\linewidth}
                \centering
                \includegraphics[width=\linewidth]{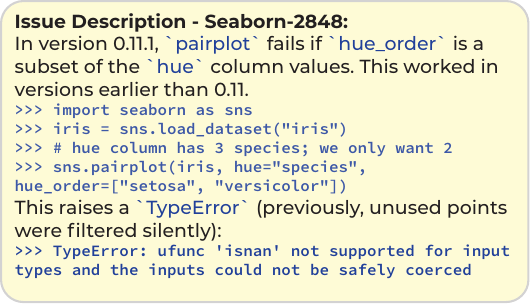}
                \label{fig:issue_desc}
            \end{subfigure}
            \vspace{-20pt}
            
            \begin{subfigure}[b]{\linewidth}
                \centering
                \includegraphics[width=\linewidth]{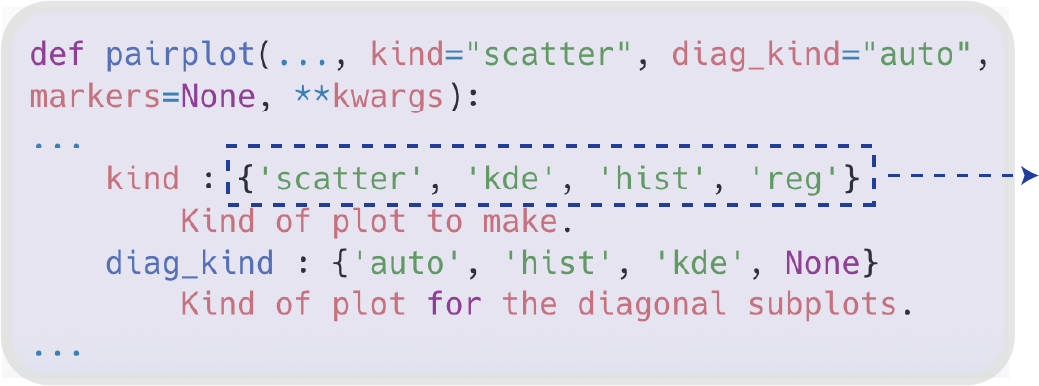}
                \caption{Add. Context: Implementation of \texttt{pairplot} in repo}
                \label{fig:pairplot_impl}
            \end{subfigure}
            
        \end{minipage}
        \hfill
        \begin{minipage}[b]{0.49\linewidth}
            \centering
            
            \begin{subfigure}[b]{\linewidth}
                \centering
                \includegraphics[width=\linewidth]{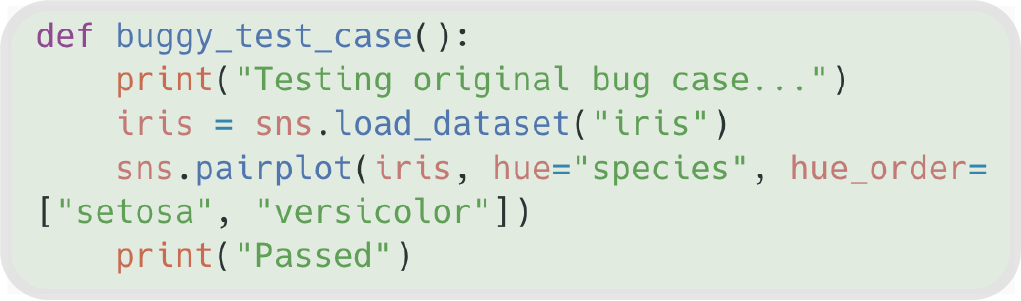}
                \vspace{-16pt}
                \caption{Reproduction Script}
                \label{fig:buggy_test}
            \end{subfigure}
            
            \begin{subfigure}[b]{\linewidth}
                \centering
                \includegraphics[width=\linewidth]{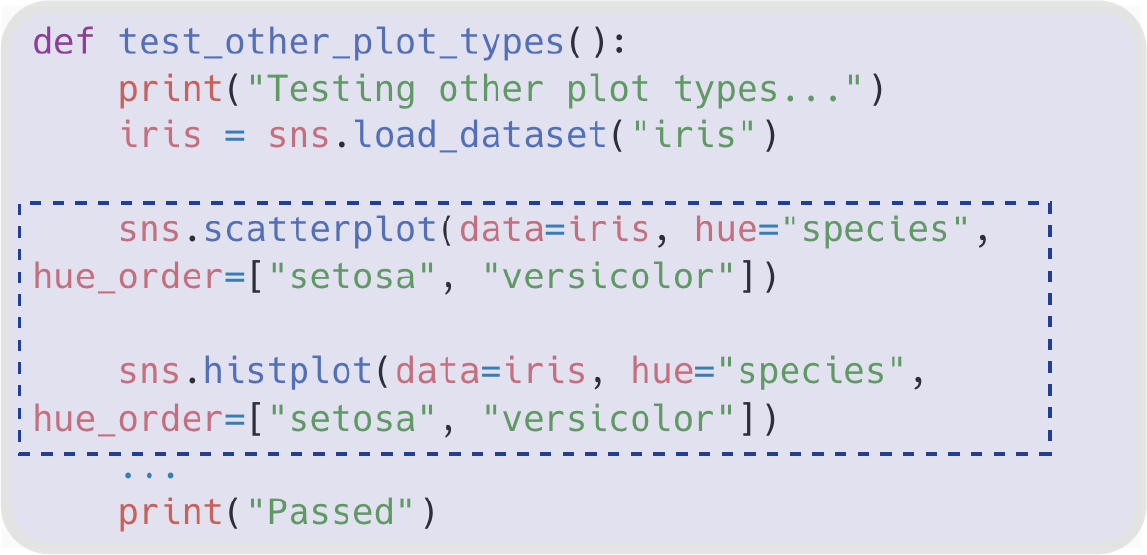}
                \vspace{-16pt}
                \caption{Additional validation test for plot types}
                \label{fig:validation_test}
            \end{subfigure}
            
        \end{minipage}

    \end{minipage}
    \caption{Seaborn-2848 Issue.}
    \label{fig:seaborn}
\end{figure}
\label{sec:example2} Fig \ref{fig:seaborn} presents another sample issue from the SWE-bench dataset, \textit{seaborn-2848}, which is a regression bug in the Seaborn visualization library. The bug occurs when drawing \texttt{pairplot}: users provide a dataset containing multiple categories but configure the plot to display only a subset of them. In previous versions, the library handled the unselected categories by rendering them as transparent. However, in the buggy version, the program crashed with an uncaught exception. The \textbf{correct behavior} is to map these unselected categories to a transparent color code (0, 0, 0, 0) rather than treating the lookup failure as a data type error. The fix involves modifying the \texttt{HueMapping.\_lookup\_single} function in the \texttt{\_oldcore.py} file.

\textbf{Existing approaches.} Leading agentic approaches \cite{refact,yang2024sweagent,experepair,kgcompass} primarily use the issue description to generate validation tests to assess the correctness of their patches. Such a validation method could lead to the \emph{symptom-fix trap}, where agents address the surface failure without resolving the underlying issue. In this example, the issue explicitly reports a failure in \texttt{pairplot}. As a result, existing approaches \cite{refact,yang2024sweagent,experepair,kgcompass} interpret the task narrowly as ``prevent \texttt{pairplot} from crashing.'' 
Concretely, they only focus on the internal function \texttt{\_plot\_bivariate}, which is the core engine behind \texttt{pairplot}.
Hence, their validation tests are limited to verifying whether the reported \texttt{pairplot} failure is resolved.
Consequently, these patches fail human-written validation tests, which also evaluate the behavior of \texttt{scatterplot}, a related plot type.

As shown in the relevant code context in Fig~\ref{fig:pairplot_impl}, \texttt{pairplot} is not an isolated component: other plotting functions (i.e., \texttt{scatterplot} and \texttt{histplot}) depend on a shared logic.
This suggests that if \texttt{pairplot} exhibits a given failure, validation should also test whether similar issues arise in these related plots.
This supports our intuition that we should \textbf{consider the broader context of the program to find relevant information that is needed to validate.} 


\finding{Key Idea 2: Context-Aware Validation Test.}{Issue descriptions often offer only a limited view of the problem, leading to generated test cases that address specific symptoms rather than the underlying issue. To overcome this, leveraging program context when generating validation tests will enhance the coverage, ensuring that edge cases and similar functions are properly tested and that the fix generalizes beyond the initial symptom.
}


\section{XAgent: eXecution-guided Agentic AI}
\label{section:method}

\subsection{Overview}



In this paper, we introduce \tool, an execution-guided agentic framework for GitHub issue resolution that analyzes dynamic behavior during program execution for localization and leverages program context to generate more reliable validation tests. 
Based on \textbf{Key Idea 1} (Section \ref{sec:example1}), we introduce the \textbf{execution-guided localizer} that compares the failing and passing execution of the program to identify functions that potentially contain the issue.
This directly addresses \textbf{Limitation 1}, overcoming the constraints of static issue descriptions that often fail to surface the deeper context needed for accurate bug localization, as demonstrated in \textbf{Example 1}.
%
To extend validation coverage beyond the issue description addressing \textbf{Limitation 2}, we develop the \textbf{context-aware validation test augmentor}.
Based on \textbf{Key Idea 2} (Section \ref{sec:example2}), our validation test augmentor uses additional contextual information from related files and functions to generate comprehensive validation tests, improving code coverage, as demonstrated in \textbf{Example 2}.

\begin{figure}[t]
    \centering
    \includegraphics[width=\linewidth]{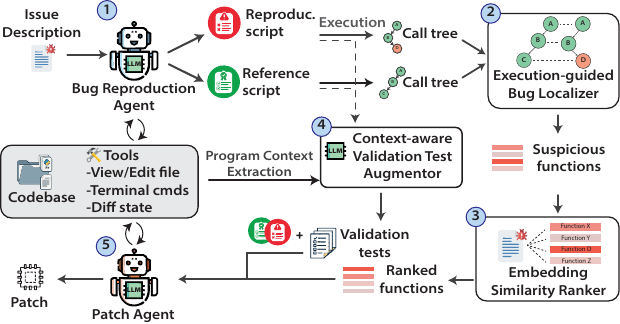}
    \vspace{-16pt}
    \caption{An overview of \tool method.}
    \label{fig:overview}
\end{figure}

Fig \ref{fig:overview} illustrates the overall workflow of \tool, which consists of three phases: reproduction, localization, and patch \& validation.
In the \textbf{reproduction phase \circled{1}}, the issue description is provided to the reproduce agent, which is responsible for generating two scripts: one that captures the buggy behavior and another that represents the non-buggy behavior (see examples in Fig. \ref{fig:buggy_trace} and \ref{fig:non-buggy_trace}).

After generating these scripts, the agent validates them by comparing their outputs against the expected results described in the reported issue. If the issue is confirmed, \tool{} will proceed the next phase. 
In the \textbf{localization phase \circled{2}-\circled{3}}, the reproduction scripts are executed to collect dynamic execution traces, which are represented as call trees (see examples in Fig. \ref{fig:buggy_trace} and \ref{fig:non-buggy_trace}). 
These two call trees are then analyzed using a tree edit distance algorithm \cite{APTED1} to identify suspicious functions that uniquely appear in the failing trace. 
The identified functions are then ranked by the embedding similarity ranker \circled{3} based on the issue description and function names to prioritize the most likely buggy functions.
The process ends with the \textbf{patch generation \& validation phase \circled{4}-\circled{5}}. In this phase, the Context-aware Validation Test Augmentor \circled{4} generates additional validation tests based on the context of related files and functions, enabling a more thorough evaluation of the bug. The set of validation tests and the ranked list of functions are then passed to the patch agent \circled{5}. This agent will iteratively identify the final buggy function, create a candidate patch, and validate it until the generated patch passes all the validation tests. 

\subsection{Reproduction}


During this phase, the reported issue description is passed to the \textbf{reproduction agent}, which generates a reproduction script and a reference script.  
The reproduction script is designed to trigger the unexpected behavior (buggy) described in the issue report, thereby confirming the presence of the bug.  
The reference script, in contrast, demonstrates the intended correct behavior (non-buggy) of the target feature, showing that the program executes as expected without errors.  
Together, these scripts establish a baseline of correct functionality and a concrete example of failure.

To guide the process, \tool{} explicitly instructs the reproduction agent to first carefully read and interpret the issue description. The agent then constructs the reproduction script to capture the erroneous behavior described in the issue. After that, it generates the reference script with a structure similar to the reproduction script but introduces minimal input differences to reflect the intended correct behavior of the feature. These small variations are essential, as large differences may introduce unrelated factors that affect execution traces and add noise, making it harder to identify the root cause during call tree comparison.  
By enforcing this similarity constraint, the agent produces clean and comparable scripts that improve the reliability of downstream analysis.

Once both scripts are generated, the reproduction agent executes them and validates their outputs against the behaviors described in the issue report. If the reference script runs successfully and the reproduction script exhibits the expected erroneous behavior, the reproduction is considered successful. Otherwise, the agent regenerates the scripts while incorporating the current results.  
Finally, both scripts are submitted to the localization phase of the pipeline.
\vspace{-5pt}

\subsection{Localization}


Once the reproduction agent submits its generated scripts, \tool{} performs localization to identify the potential buggy functions. 
Particularly, \tool{} executes the two scripts using a tracing module, which records function calls, including the caller--callee relationships and any associated traceback errors. From this execution trace, we construct call trees, where each node corresponds to a function, and each child node represents a callee of its parent.

Let $T_1 = (V_1, E_1)$ and $T_2 = (V_2, E_2)$ represent the call trees for the buggy and non-buggy executions, respectively, where $V_1$ and $V_2$ are the sets of nodes (functions) in $T_1$ and $T_2$, and $E_1 \subseteq V_1 \times V_1$ and $E_2 \subseteq V_2 \times V_2$ are the edge sets representing caller--callee relationships.


To capture the behavioral differences between the two call trees, we apply the All Path Tree Edit Distance (\textbf{APTED}) algorithm \cite{APTED1,apted2,apted3}.
APTED is an efficient algorithm for computing the edit distance between two trees. It produces a set of node mappings that explain how one tree can be transformed into the other, while minimizing the number of edit actions (i.e., delete, insert). The output mappings fall into three categories.
Given the tree edits from a buggy tree to a non-buggy tree ($T_1 \rightarrow T_2$) and the node mappings \( (u, v) \in M \) where \( u \in V_1 \) and \( v \in V_2 \), each category is defined as below:

\begin{itemize}[leftmargin=*]
\item{ \emph{Identical}: $u$ and $v$ are matched without any modification.}

\item{ \emph{Deletions}: \( (u, \epsilon) \in M \), node \(u\) is deleted from the buggy tree.}

\item{ \emph{Insertions}:  \((\epsilon, v) \in M \) where  node \(v\) is inserted to the buggy tree.}
\end{itemize}

For localization, we focus on nodes (functions) that appear uniquely in the buggy execution (\emph{deleted nodes}). We denote them as the potential buggy functions \emph{\textbf{ \(S = \{\text{deleted nodes}\}.\)}}    
Additionally, we incorporate knowledge from the issue description and the reproduction phase. Specifically, functions explicitly mentioned in the issue report or highlighted by the reproduction agent are often highly relevant to the bug. If such functions are not included in \( S \), we add them to ensure they are included in downstream analysis.

\textbf{Embedding Similarity Ranker.} To improve efficiency and prioritize the most suspicious candidate functions, we rank functions in \( S \) by their similarity to the issue description using UniXcoder \cite{guo2022unixcoderunifiedcrossmodalpretraining}. We encode the issue description, each function body, and each function name with the UniXcoder tokenizer and encoder, then obtain a single vector for each by mean-pooling the last hidden state and applying L2 normalization. We compute cosine similarities between the issue vector and the vectors of the function body and name to derive a relevance score, sort \( S \) by this score to obtain \( S' \), and retain the top \hieu{$K$} functions for subsequent steps.

\subsection{Patch Generation \& Validation}
After localizing the potential buggy functions, \tool{} proceeds to the final phase of patch generation and validation. 
To \pick{help the agent} more thoroughly evaluate the \pick{generated} patches, we extend the validation beyond the two reproduction \pick{scripts} by generating additional validation tests as motivated in Section \ref{sec:example2}.
To this end, we introduce \textit{the context-aware \pick{validation} augmentor} in \tool, which generates validation tests based on the program context beyond the code context described in the issue report.
By doing so, they help ensure that the patch agent not only fixes the reported bug but also maintains overall functionality and correctness across related components.


\textbf{Context-aware Validation Augmentor.} To generate additional validation tests, the \pick{validation} augmentor leverages program context extracted directly from the reproduction agent's trajectory.


Broadly speaking, an agent trajectory is a structured record of the agent’s interaction with the environment, consisting of iterative \emph{thought–action–\\observation} tuples (e.g., reasoning traces, tool invocations, and returned outputs). For our reproduction agent, the trajectory captures how the agent navigates the repository, inspects code snippets, and how different files and functions relate to the issue description. This process surfaces a filtered set of context—i.e., files and functions already deemed relevant by the agent’s problem-solving process—providing a strong foundation for validation. We then leverage this trajectory as task-specific context to generate additional validation tests through two steps.
First, we prompt an LLM to analyze the recorded trajectory to extract \emph{program context}—specifically, the files and functions the reproduction agent found relevant to the issue description.
Then, we prompt an LLM to generate an additional validation script based on (1) the extracted program context, (2) the original reproduction scripts, and (3) the issue description.
As all relevant context has been obtained, there is no need to use tools to extract additional program context, so we use zero-shot prompting rather than an iterative, tool-enabled agent.
We reuse the context gathered during reproduction rather than letting the validation augmentor independently navigate the codebase, which eliminates redundant exploration and substantially reduces both token consumption and computational cost.



\pick{
\textbf{Patch Generation.} Finally, \tool{} generates the patch based on the ranked list of functions \( S' \) identified by our execution-guided localizer and the issue description based on the following three steps:
First, the patch agent executes our five validation tests (i.e., three context-aware validation and two reproduction scripts) and existing regression tests to ensure that the environment is correctly set up and that the bug is reproducible. 
Second, the agent is guided to iterate through each ranked function in \( S' \). For each function, the agent determines whether it is related to the issue. If so, it attempts to generate a candidate patch; otherwise, it moves on to the next function in \( S' \). 
Third, once a patch candidate is generated, the agent reruns our five validation tests to check that the fix resolves the issue as well as the regression tests to confirm that the fix does not adversely affect existing functionality.  If the patch fails to address the issue, the agent receives feedback from the environment and attempts to regenerate the patch. This process continues until either the generated patch successfully passes the validation and regression tests or a predefined limit is reached.
}

\section{Experimental Setup}
\label{section:setup}
\subsection{Research Questions}
To evaluate \tool, we formulate the following three RQs:
\begin{enumerate}[label=\textbf{(RQ\arabic*}),leftmargin=*]
    \item \textbf{Effectiveness and Efficiency:} How accurate and cost-efficient is \tool{} in bug localization and issue resolution?
    
    \item \textbf{Ablation Study:} How does each key component of \tool{} contribute to its overall performance?
    
    \item \textbf{Resolve Rate vs. Issue Complexity:} How does the resolve rate change across different levels of issue complexity?
\end{enumerate}

\subsection{Benchmark} To address our research questions, we adopt the widely used benchmark \textit{SWE-bench-Lite} \cite{yang2024swebench}, specifically designed to evaluate the ability of LLMs to resolve real-world software engineering tasks at the repository level. Unlike traditional code generation benchmarks such as HumanEval \cite{humaneval} or MBPP \cite{mbpp}, which focus on isolated functions, issues in \textit{SWE-bench-lite} involve repository-level challenges that require understanding of code dependencies across multiple files and comprehensive reasoning about software behavior.
This benchmark provides a publicly available leaderboard, enabling direct comparison with other state-of-the-art methods. We select this dataset to ensure consistency with prior studies \cite{yu2025orcaloca,yang2024sweagent,jiang2025cosil,experepair} and to maintain computational efficiency, as it already includes standardized evaluation results.
 \textit{SWE-bench-Lite} consists of 300 real GitHub issues (including bug reports and feature requests) drawn from 12 popular open-source Python repositories. Each instance includes an issue description and the corresponding repository, where the goal is to modify the relevant source code to resolve the issue without prior knowledge of the issue location. 
\vspace{-7pt}

\subsection{Experimental Setup} 
In this experiment, we used \texttt{Claude Sonnet 4} as the backbone LLM because of its strong coding capabilities and competitive performance on SWE-bench tasks. 
This is also consistent with the top method (i.e., ExpeRepair), making our choice consistent with prior SOTA methods and ensuring fairness in comparison.
\pick{We limit the number of suspicious functions to $K=20$, as this threshold covers 80\% of cases in our experiments (Table \ref{tab:ablation_ranking}) while remaining manageable for the LLM to search.}
As the reproduction and patch agents autonomously perform actions (i.e., view and edit files, execute terminal commands, and inspect diffs after modifications), we set the action limits to control the experimental costs.
Particularly, action limits are 100 for the reproduction agent and 200 for the patch agent. 
\pick{
For the validation augmentor, we set the temperature parameter to 0.4, 0.6, and 0.8 to generate three varied validation tests to increase the variation in generation.
For the other tasks, we configured the temperature parameter to match their task requirements: $0$ for reproduction to ensure consistency and adherence to the issue description, and $0.5$ for patch generation to maintain a balance between creative problem-solving and code accuracy \cite{temperature}.}

\vspace{-7pt}

\subsection{Baselines}
\label{sec:baseline}
 We select nine methods representing the top-performing methods on the SWE-bench-lite leaderboard: ExpeRepair \cite{experepair} (Claude 4 Sonnet), Refact Agent \cite{refact}, SWE-Agent \cite{yang2024sweagent} (Claude 4 Sonnet), DARS \cite{aggarwal2025dars}, KGCompass \cite{kgcompass}, CodeFuse \cite{codefuse}, OpenHands \cite{wang2024openhands}, Composio \cite{composio}, and OrcaLoca \cite{yu2025orcaloca}. These methods hold the highest rankings on the leaderboard. Each method has officially submitted its results, and we reuse their public leaderboard submissions for comparison.

\vspace{-7pt}
 
\section{Experimental Results}
\label{section:experimental_results}

\subsection{RQ1: How accurate and cost-efficient is \tool{} in bug localization and issue resolution?}
\label{section:rq1}
\smallsection{Approach}
To address this RQ, we compare \tool{} against 9 state-of-the-art baselines introduced in Section \ref{sec:baseline} on 300 test instances from the SWE-bench-lite benchmark.

To evaluate the \textbf{bug localization} of our \tool{}, we use \textbf{Precision}, \textbf{Recall}, and \textbf{F1-score}, which provide a fine-grained assessment of how accurately an approach identifies true buggy locations while penalizing unnecessary edits. Specifically, let $S$ denote the set of files or functions modified in the ground-truth patch, and $S'$ those modified by the approach. We define $TP = |S' \cap S|$, $FP = |S' \setminus S|$, and $FN = |S \setminus S'|$. The corresponding metrics are computed as $P = \frac{TP}{TP + FP}$, $R = \frac{TP}{TP + FN}$, and $F1 = \frac{2PR}{P + R}$. These metrics are applied at both the file and function levels.
An F1 score of 1 indicates that the predicted buggy files (or functions) exactly match those modified in the golden patch, reflecting a perfect bug localization.
This formulation addresses limitations of prior evaluations \cite{agentless,yu2025orcaloca}, which rely on a superset match criterion that deems a localization correct if all ground-truth locations are edited, even when many irrelevant files or functions are also modified. Such a criterion can overestimate performance by rewarding high recall despite low precision, or penalize minimal yet correct fixes that omit unnecessary edits. In contrast, our formulation explicitly captures the trade-off between localization precision and edit redundancy.

To evaluate the \textbf{issue resolution} of our \tool{}, we follow prior works \cite{yang2024sweagent,zhang2024autocoderover,experepair} and use the \textbf{\%Resolved} metric, which measures the proportion of GitHub issues that an approach can successfully resolve. Specifically, a software issue is considered resolved if the generated patch (i) can be correctly applied to the target codebase, and (ii) passes all associated unit tests, especially the \textit{fail-to-pass} tests that verify the original defect has been fixed. These unit tests are written by human developers to ensure correctness and are kept hidden from the model to prevent test data leakage.

To evaluate the \textbf{cost-efficiency} of our \tool{} for both bug localization and issue resolution tasks, we measure both LLM API cost and token usage. For baseline methods, we report the total monetary cost (in USD) as stated in their original papers; when such information is unavailable, it is marked as “–”.
Token usage includes all input and output tokens exchanged with the LLM, including system prompts, user inputs, model responses, and tool calls. For baselines, token counts are recomputed from their released trajectories to ensure fair comparison.

\begin{figure}[t]
    \centering
    \includegraphics[width=0.6\linewidth]{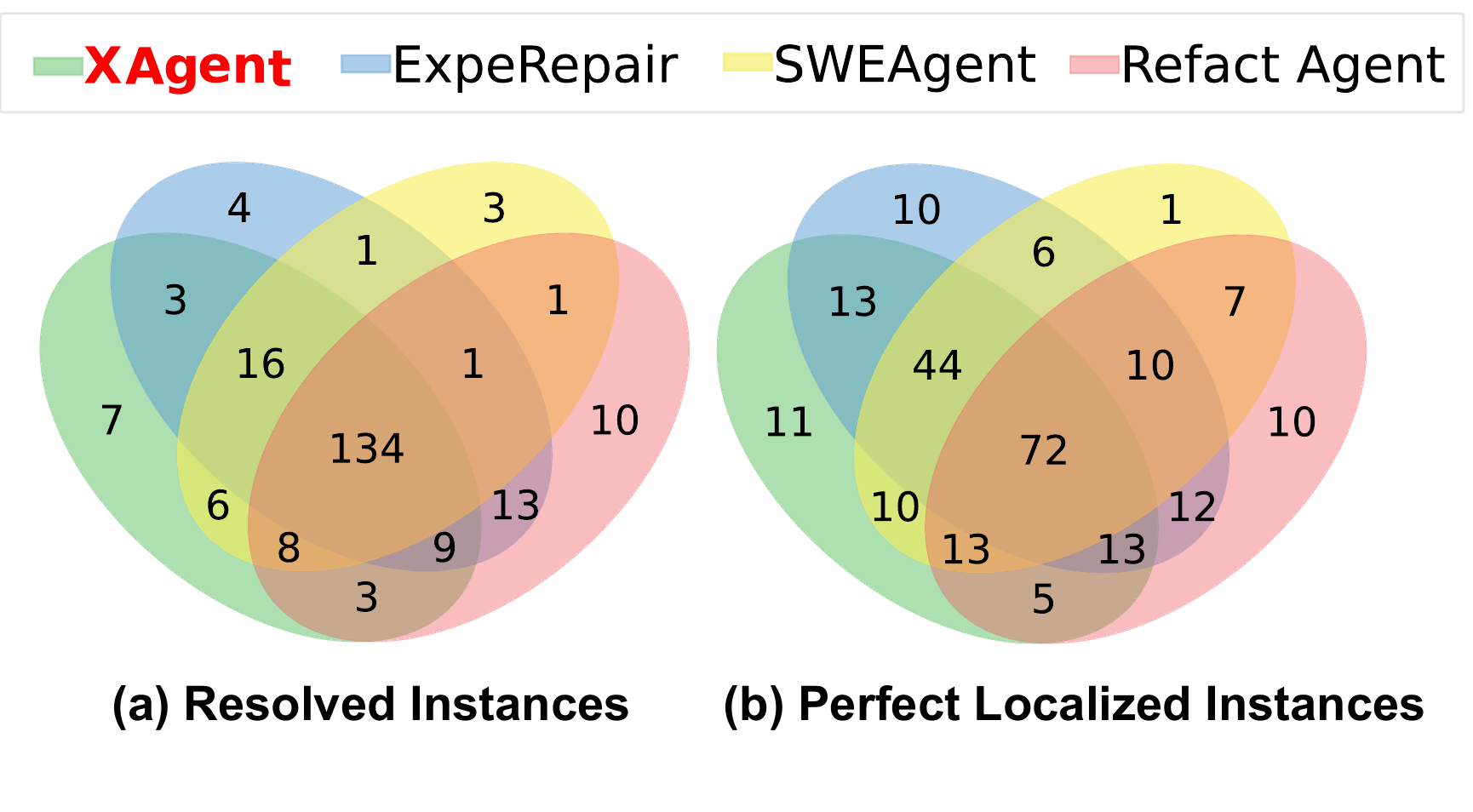}
    \caption{Overlapping Analysis.}
    \label{fig:venns}
\end{figure}

\begin{table}[t]
\centering
\setlength{\tabcolsep}{3pt}
\caption{Performance on SWE-bench-lite dataset evaluation. The best results for each metric are \textbf{bolded}.}
\label{tab:overview_results}
    \resizebox{\columnwidth}{!}{%
            \begin{tabular}{l|r|r|r|rrr|rrr}
            \toprule
            \multirow{2}{*}{\textbf{Method}} & \multirow{2}{*}{\textbf{\%Resolve}} & \multirow{2}{*}{\textbf{Cost}} & \multirow{2}{*}{\textbf{Token}} & \multicolumn{3}{c|}{\textbf{File}} & \multicolumn{3}{c}{\textbf{Function}} \\
\cmidrule(lr){5-7} \cmidrule(lr){8-10}
 &  &  &  & \textbf{P} & \textbf{R} & \textbf{F1} & \textbf{P} & \textbf{R} & \textbf{F1} \\

\midrule
\textbf{XAgent (\claudelogo-$4$)} & \textbf{\cellcolor{green!40}{62.0 (186)}} & 1.56 & 426.5k & \textbf{\cellcolor{green!40}{85.1}} & \cellcolor{green!25}{89.7} & \textbf{\cellcolor{green!40}{86.6}} & \textbf{\cellcolor{green!40}{72.6}} & \cellcolor{green!25}{76.9} & \textbf{\cellcolor{green!40}{72.8}} \\
\midrule
ExpeRepair (\claudelogo-$4$\&o$4$m) & \cellcolor{green!25}{60.3 (181)} & 2.49 & 713.3k & \cellcolor{green!25}{84.7} & \cellcolor{green!5}{88.0} & \cellcolor{green!25}{85.7} & \cellcolor{green!25}{70.3} & \cellcolor{green!5}{74.8} & \cellcolor{green!25}{70.7} \\
Refact (\claudelogo-$3.7$\&o$4$m) & \cellcolor{green!15}{60.0 (180)} & - & - & 78.3 & \textbf{\cellcolor{green!40}{93.6}} & \cellcolor{green!15}{83.0} & 64.6 & \textbf{\cellcolor{green!40}{77.9}} & \cellcolor{green!5}{67.7} \\
SWE-Agent (\claudelogo-$4$) & \cellcolor{green!5}{56.7 (170)} & 1.18 & 379.3k & 78.6 & \cellcolor{green!15}{89.6} & 81.5 & \cellcolor{green!15}{67.6} & \cellcolor{green!15}{76.5} & \cellcolor{green!15}{69.0} \\
KGCompass (\claudelogo-$3.5$) & 46.0 (138) & 0.20 & - & 79.0 & 79.0 & 79.0 & 65.5 & 62.3 & 63.3 \\
CodeFuse (CGM) & 44.0 (132) & - & - & 60.7 & 63.3 & 61.6 & 28.0 & 28.5 & 27.7 \\
OpenHands (\claudelogo-$3.5$) & 41.7 (125) & - & - & 45.4 & 82.6 & 54.4 & 36.3 & 62.8 & 42.9 \\
Composio (\claudelogo-$3.5$\&o1) & 41.0 (123) & - & - & 78.4 & 80.3 & 79.0 & 62.5 & 63.3 & 61.8 \\
OrcaLoca (\claudelogo-$3.5$) & 41.0 (123) & 1.77 & - & \cellcolor{green!5}{80.6} & 80.7 & 80.6 & \cellcolor{green!5}{66.9} & 64.0 & 64.8 \\
Moatless (\claudelogo-$3.5$) & 39.0 (117) & - & - & 59.0 & 82.4 & 66.3 & 46.3 & 63.2 & 51.2 \\
            \bottomrule
            \end{tabular}
}
\begin{minipage}{\linewidth}
\footnotesize
\claudelogo: Claude Sonnet models. o4m: GPT-o4-mini. \raisebox{0.5ex}{\colorbox{green!40}{}} \raisebox{0.5ex}{\colorbox{green!25}{}} \raisebox{0.5ex}{\colorbox{green!15}{}} \raisebox{0.5ex}{\colorbox{green!5}{}}: Rank 1--4.
\end{minipage}
\end{table}
\smallsection{Results}
Table \ref{tab:overview_results} presents the bug localization and issue resolution results of \tool and 9 other methods on the SWE-bench-lite benchmark across different metrics.

\textbf{Our \tool{} achieves the best overall performance across both tasks, with file- and function-level localization F1 scores of 86.6\% and 72.8\%, and the highest issue resolution rate of 62\%.}
In particular, \tool{} reduces token usage by nearly 40\% compared to the prior state of the art, resulting in an overall cost reduction of approximately \$300 across all instances. This improvement is primarily due to avoiding LLM calls during localization and eliminating the need for generating and selecting multiple candidate patches, which together save about \$1 per instance. \textbf{These results demonstrate that \tool{} achieves higher repair accuracy while being substantially more cost-efficient than the existing best approach.}

\textbf{In terms of bug localization}, \tool{} achieves the best performance across all key metrics, with file-level Precision and F1 scores of 85.1\% and 86.6\%, and function-level Precision and F1 scores of 72.6\% and 72.8\%, respectively. While Refact Agent achieves the highest recall at both file and function levels, it achieves substantially lower precision, indicating a tendency to modify many non-buggy locations. This behavior reflects a common pattern across baseline methods, which exhibit higher recall than precision, suggesting that they often over-approximate the buggy region by editing more files or functions than necessary. In contrast, \tool{} achieves a more balanced trade-off, maintaining high recall while significantly improving precision, thereby reducing redundant~edits.



Fig \ref{fig:venns}(b) presents an overlap analysis between our \tool{} and the three top-performing baselines—ExpeRepair, Refact Agent, and SWE-Agent—based on perfect localization, defined as instances achieving an F1 score of 100\%. 
Our \tool{} correctly localizes the largest number of instances (181) and uniquely identifies the correct buggy locations in 11 cases that none of the other methods handle correctly. In comparison, Refact Agent, ExpeRepair, and SWE-Agent achieve 10, 10, and 1 unique instances, respectively. We further analyzed these 11 cases and found that most successes (7/11) stem from our differential analysis, which effectively captures behavioral differences between failing and passing executions. The remaining cases benefit from our \tool's exploratory reasoning during the reproduction phase. Although Refact Agent identifies 10 unique buggy locations, it achieves the lowest total number of perfectly localized instances (142) among the top methods, largely due to its tendency to modify more files or functions than necessary. \textbf{Overall, these results demonstrate that \tool{} provides more accurate and reliable bug localization at both the file and function levels than existing approaches.}

The improved localization accuracy also translates into better issue resolution performance. As a result, our \tool{} resolves 186 issues, outperforming ExpeRepair, Refact Agent, and SWE-Agent by 5, 6, and 16 instances, respectively. Fig \ref{fig:venns}(a) presents an overlap analysis between our \tool{} and the three top-performing baselines on issue resolution.
We observed that other baselines often fail to localize the bug at the function level or generate fixes that only address the symptom described in the issue report.
In contrast, human developers typically identify underlying causes and implement solutions that generalize beyond specific reported~issues.

A representative example that only \tool{} is able to fix correctly is \textit{matplotlib-24265}, which describes a behavioral bug in the Matplotlib repo, where attempting to load the \texttt{seaborn-colorblind} style using \texttt{plt.style.\\library} raises a \texttt{KeyError}.
This style has been deprecated since version 3.6.1; however, instead of raising a deprecation warning, the program crashes. The correct fix involved renaming the style file to \texttt{seaborn-v0\_8-colorblind} and ensuring that accessing the old \texttt{seaborn-colorblind} key now raises a \texttt{Matplotlib\\DeprecationWarning} rather than an error.
To validate the fix, developers added a unit test to confirm that (1) the new \texttt{seaborn-v0\_8-colorblind} style loads successfully, and (2) accessing the deprecated \texttt{seaborn-colorblind} key triggers the appropriate warning. However, the three baseline methods generate patches that address only the former, neglecting the required warning behavior and thus producing incorrect fixes.

In contrast, \tool{} begins patch generation by executing both reproduction and validation tests to understand how the program behaves in failure cases and in expected usage scenarios before generating any patch. During validation, \tool{} tests both \texttt{plt.style.library[“seaborn-colorblind”]} and \texttt{plt.style.use(“seaborn-colorblind”)}, an alternative access path for the same style. Through this process, \tool{} observes that the former raises a \texttt{KeyError}, whereas the latter executes successfully and emits a deprecation warning. This discrepancy guides \tool{} to inspect the underlying code paths, revealing that the warning logic is already implemented in \texttt{plt.style.use()}. \tool{} therefore applies the same to \texttt{plt.style.library[]}, resulting in a correct patch that passes all unit tests. This example illustrates how \tool{}’s strategy of deriving diverse validation tests from reproduction cases enables it to uncover subtle behavioral inconsistencies and generate correct fixes that resolve the root cause. \textbf{Overall, these results show that \tool{} provides more accurate and reliable issue resolution by enhancing the coverage of the bug instead of addressing only the specific symptoms described in the issue report.}

\finding{Answer to RQ1.}{\tool{} achieves the best overall performance on SWE-bench-Lite, with the highest bug localization accuracy and issue resolution rate, while reducing both monetary cost and token usage compared to the prior top-performing method (i.e., ExpeRepair). This demonstrates that \tool{} provides accurate, efficient, and comprehensive issue resolution.}

\subsection{RQ2: How does each key component of \tool{} contribute to its overall performance?}
\label{section:rq2}
\smallsection{Approach}
To answer this RQ, we investigate the two key components in our \tool: execution-guided bug localizer and context-aware validation test augmentor.
Referring to the workflow illustrated in Fig~\ref{fig:overview}, we introduce two variants of \tool:
\begin{itemize}[leftmargin=*]

    \item{ \textbf{\textit{Vanilla Agentic LLM}}: A baseline agentic approach without the use of our \textit{execution-guided bug localizer} (\circled{2} and \circled{3}) and \textit{context-aware validation test augmentor} (\circled{4}). This variant follows the same design as the prior agentic approach, SWE-Agent, employing a single LLM agent to reproduce the issue, localize the bug solely based on the information provided in the issue description, and generate a patch accordingly. The generated patch is then validated using the existing test suite, without any additional test augmentation.}
    
    \item{ \textbf{\textit{Vanilla Agentic + Execution-guided Bug Localizer}}: This variant extends the \textbf{Vanilla Agentic LLM} by incorporating the \textit{execution-guided bug localizer} (\circled{2} and \circled{3}). 
    Then, the LLM agent generates a patch, which is subsequently validated using the reproduction and regression tests without any additional test augmentation. \textit{This variant isolates and quantifies the contribution of the execution-guided bug localization component.}}
    
    \item{ \textbf{\textit{Vanilla Agentic +  Execution-guided  Bug Localizer + Validation Test Augmentor} (\tool)}: This variant further incorporates the \textit{context-aware validation test augmentor} (\circled{4}) to form our \tool approach. After identifying suspicious functions, the LLM agent generates additional validation tests based on the context of the localized files and functions. It then generates a patch and validates it against both the original and the augmented tests. \textit{This variant quantifies the additional benefit of validation test augmentation beyond execution-guided localization.}}
\end{itemize}

In addition, we conduct an ablation on the \textbf{\textit{Embedding Similarity Ranker}} (\circled{3}) by altering the ranking method:
\begin{itemize}[leftmargin=*]

    \item{\textbf{\textit{Embedding Similarity Ranker}}: This ranker is used by \tool{} to prioritize candidate functions based on their semantic similarity to the issue description using UniXcoder and compute cosine similarity between their embeddings. Functions are then ranked according to this similarity score.}
    
    \item{ \textbf{\textit{Commit History Ranker}}: Prior work~\cite{hata2012bug,hoang2020cc2vec} argued that functions modified more frequently in recent history are more likely to contain bugs. Hence, we rank the candidate functions based on their change frequency over the past year, producing an ordered list used for subsequent localization.}
    
    \item{ \textbf{\textit{Proximity-based Ranker}}: Inspired by OrcaLoca~\cite{yu2025orcaloca}, this ranker leverages simple proximity-based signals derived from execution traces. Specifically, it prioritizes functions that reside in the same file as suspicious keywords, appear along descendant paths in the trace, or are closer to the root of the execution tree. These signals are combined to score and rank candidate functions for bug localization.}
\end{itemize}
We assess ranking quality using a \textbf{Top-\(K\)} metric, which measures whether the ground-truth buggy function appears within the top \(K \in \{10, 20\}\) ranked candidates.

\begin{table}[t]
\centering

\setlength{\tabcolsep}{3pt}
\caption{Contribution of Key Components.}
\label{tab:keycomponents}

\begin{tabular}{l|l|l}
\hline
\textbf{Method} & \textbf{\%Resolve (\(\Delta\))} & \textbf{Func. F1 (\(\Delta\))} \\
\hline
\textbf{Vanilla Agentic LLM} & 56.7 & 69.0 \\
~ \textbf{+ Bug Loc.} & 58.0 (\(\uparrow 1.3\)) & 73.5 (\(\uparrow 4.5\)) \\
~ \textbf{+ Bug Loc. + Val. Test (\tool)} & 62.0 (\(\uparrow 5.3\)) & 72.8 (\(\uparrow 3.8\)) \\
\hline
\end{tabular}

\end{table}

\smallsection{Results}
Table \ref{tab:keycomponents} presents the results of \tool{} and its two variants.
When incorporating our \textbf{\textit{execution-guided bug localizer}} into the vanilla agentic LLM, we observe \pick{a clear improvement in  F1-score}. Specifically, our bug localizer increases the function-level F1 score by 4.5\%, achieving the highest Function Localization F1 of 73.5\%. In addition, the \%Resolved metric improves by 1.3\%, corresponding to four more successfully resolved issues compared to the vanilla agentic LLM variant. 
\pick{We also observe that, with our execution-guided bug localizer, the agent uses 13\% fewer tokens than the vanilla agentic LLM.
Consequently, LLM usage and associated costs are reduced.} 
\textbf{These results demonstrate the effectiveness of the \textbf{\textit{execution-guided bug localizer}} in accurately identifying buggy functions.}



For \textbf{\textit{agentic LLM + bug localizer + validation test augmentor} (\tool)}, we observed the best overall performance, achieving a \%Resolved score of 62.0\%. Compared to \textbf{\textit{Agentic LLM + Bug Localizer}}, this represents an improvement from 58.0\% to 62.0\%, corresponding to 12 additional resolved issues attributable to the introduction of our validation test augmentor. The slight decrease of function-level F1 score (by 0.7\%) is due to the inclusion of additional validation tests that encourage broader code modifications, mildly reducing precision. \textbf{Overall, these results demonstrate that our \textbf{\textit{validation test augmentor}} plays a critical role in improving issue resolution by evaluating patches against a richer set of test cases, enabling more generalizable fixes.}

\begin{table}
\centering

\caption{Ablation of Ranking Methods.}

\label{tab:ablation_ranking}
\begin{tabular}{lrr}
\hline
\textbf{Ranker} & \textbf{Top-10} & \textbf{Top-20} \\ \hline
\textbf{Embed. Sim. (\tool)} & \textbf{75.7} & \textbf{79.0} \\ 
\textbf{Commit History} & 73.3 & 73.3 \\ 
\textbf{Proximity-based} & 74.3 & 76.3 \\ \hline
\end{tabular}
\end{table}

In terms of ranking strategies, Table~\ref{tab:ablation_ranking} presents the Top-10 and Top-20 localization performance of the three ranking methods: \textbf{\textit{Embedding Similarity Ranker}}, \textbf{\textit{Commit History Ranker}}, and \textbf{\textit{Proximity-based Ranker}}. 
The embedding similarity ranker used in \tool{} achieves the highest Top-10 and Top-20 accuracy.
Despite the common intuition of past tendency~\cite{hata2012bug,hoang2020cc2vec}, the commit history ranker often fails when issues originate from older or long-stable code. As a result, it cannot reliably capture the semantic intent or reasoning reflected in issue descriptions. The proximity-based ranker, on the other hand, prioritizes functions closer to the root of the execution trace. However, in many cases, the true fault lies deeper in the call chain, causing this proximity-based heuristic to misrank relevant functions.
\textbf{Overall, these results indicate that semantic similarity between candidate functions and the issue description provides a reliable signal for ranking functions in \tool{}.}

\finding{Answer to RQ2.}{Our ablation study shows that each component of \tool{} contributes to performance improvement. Execution-guided localization improves localization accuracy and reduces cost. Validation test augmentor increases issue resolution rates and semantic similarity provides effective ranking. These components collectively enable \tool{} to achieve strong performance in both accuracy and efficiency for automated issue resolution.}


\subsection{RQ3: How does the resolve rate change across different levels of issue complexity?}
\label{section:taskcomplexity}
\smallsection{Approach}
To address this RQ, we analyze the issue resolution capability of our \tool approach and the three top-performing baselines (ExpeRepair, Refact Agent, and SWE-Agent) with different levels of issue complexity.
Following prior work that characterizes bug complexity by the structure and effort required for repair~\cite{bohme2014corebench,xin2024detecting}, we define issue complexity by the amount of human effort needed to fix the issues, using three metrics: (1) how many code hunks are produced in the ground-truth (human-written) patch (\#Hunks), (2) how many functions are modified (\#Functions), and (3) how many lines are removed and added (\#Lines).
Larger values in these metrics indicate higher issue complexity, as they reflect greater human effort required for fixing the bug.
We then report the \%Resolved rate across different levels of issue complexity to analyze how resolution performance varies as complexity increases.
\begin{figure}[t]
    \centering
    \includegraphics[width=\linewidth]{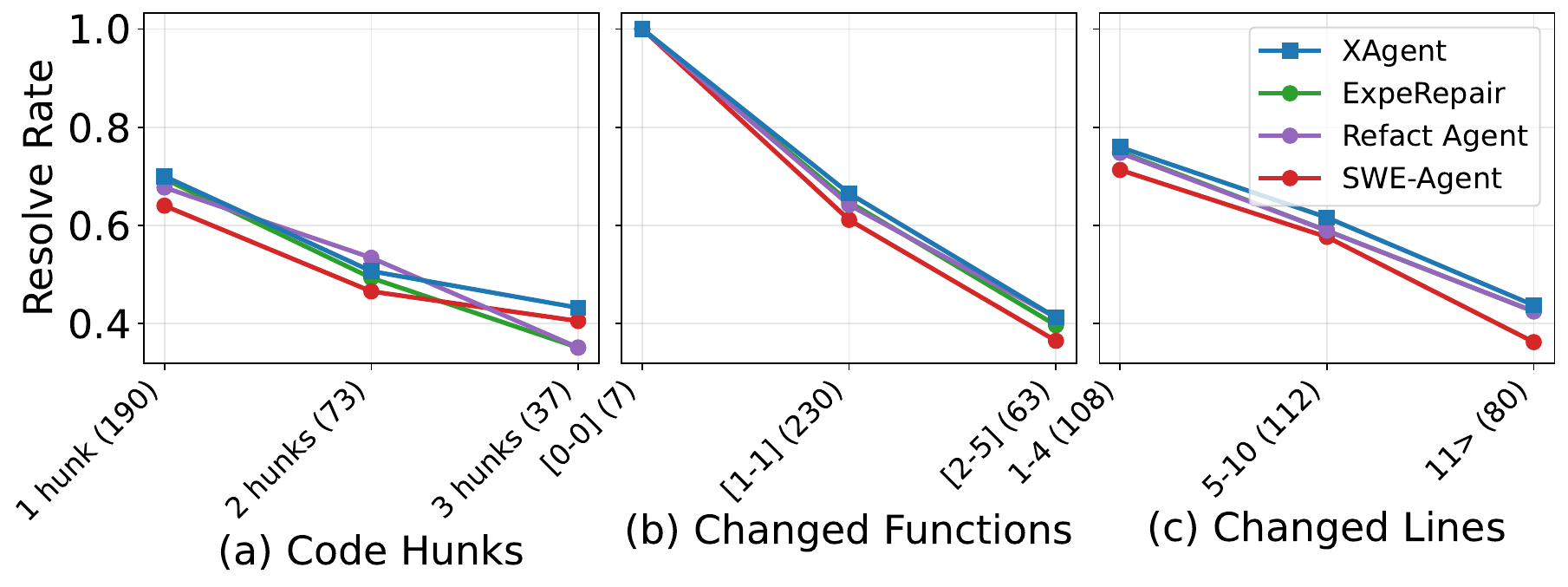}
    \vspace{-16pt}
    \caption{Resolve Rate across different complexity metrics.}
    \label{fig:all_difficulties}
\end{figure}


\smallsection{Results}
Fig~\ref{fig:all_difficulties} presents the relationship between the \%Resolve and the complexity of instances on the three metrics (\#Hunks, \#Functions, and \#Lines).

Fig~\ref{fig:all_difficulties}(a) illustrates the relationship between \%Resolve and the number of hunks (\#Hunks) in patches.
The majority of instances (190 out of 300) contain a single hunk, followed by 73 instances with two hunks and 37 instances with three hunks. All evaluated methods perform best on single-hunk cases, where our \tool{} achieves the highest \%Resolved rate at 70\%, while SWE-Agent performs the worst at 64\%. For those instances requiring two hunks, Refact Agent attains the highest resolution rate at 54\%, followed closely by our \tool at 51\%, with SWE-Agent again ranking lowest at 47\%. For more complex cases involving three hunks, our \tool again achieves the best performance with a \%Resolved rate of 44\%, whereas both ExpeRepair and Refact Agent drop to 35\%.
\textbf{These results confirm that \tool{} remains effective when resolving issues that require changes across multiple code hunks.}

Fig~\ref{fig:all_difficulties}(b) presents the relationship between the \%Resolved and the number of modified functions (\#Functions). We categorize the 300 instances into three groups: no function modified, one function modified, and two to five functions modified. Seven instances involve no function-level changes, as the fixes affect only global configurations (e.g., imports or global variables); all four studied methods successfully resolve these cases.
Among the 230 instances requiring changes to a single function, our \tool approach achieves the highest \%Resolved at 67\%, followed by Refact Agent, ExpeRepair, and SWE-Agent. For the remaining 63 instances that involve modifications to two to five functions, our \tool approach again achieves the highest \%Resolved rate at 41\%, tying with Refact Agent and outperforming ExpeRepair and SWE-Agent.
\textbf{These results confirm that \tool{} remains effective when resolving issues that require modifications across multiple functions.}

Finally, Fig~\ref{fig:all_difficulties}(c) illustrates the relationship between the \%Resolved and the number of changed lines (\#Lines).
We found that as the number of modified lines increases, the \%Resolved steadily decreases. Across all ranges of line changes. Nevertheless, our \tool{} approach consistently achieves the highest resolution performance.
\textbf{These results demonstrate that \tool{} scales effectively to fixes requiring changes to a larger number of lines of code.}

\finding{Answer to RQ3.}{As issue complexity increases, issue resolution performance generally declines across all methods. However, in the most complex cases—larger numbers of hunks, modified functions, and lines—\tool{} consistently achieves the highest resolve rate among the evaluated approaches, indicating stronger effectiveness in challenging issue~resolution.}
\section{Discussion}
In this section, we further discuss the performance of \tool{}.



\subsection{How much search space can be reduced by our execution-guided bug localizer?} 

\begin{table}[t]

  \centering
  \caption{Number of functions per issue.}
    \vspace{-5pt}
  \label{tab:functions_stats}
  \begin{tabular}{lrrrr}
    \hline
    \textbf{Strategy} & \textbf{Min} & \textbf{Max} & \textbf{Avg} & \textbf{Med} \\ \hline
    \textbf{The whole repo}      & 702    & 33,618    & 20,769  & 5,946   \\ \hline
    \textbf{buggy scenario only} & 1      & 1,049     & 234     & 128     \\
    \textbf{$\Delta$ buggy \& non-buggy (ours)}  & 1      &  637 ($\downarrow$39\%)&  52 ($\downarrow$78\%)&  11 ($\downarrow$91\%)\\ \hline
  \end{tabular}
\end{table}

As discussed in Section \ref{sec:example1}, comparing buggy and non-buggy executions could reduce the search space of buggy functions.
Hence, we further analyze the ability of \tool in reducing the function search space.  
Table~\ref{tab:functions_stats} presents the size of the search space based on 3 strategies: the whole repository, the functions that are executed in the buggy scenario, and the functions that \textbf{uniquely} executed in the buggy scenario, but not in the non-buggy scenario. 
By searching the whole repository, the search space for identifying buggy functions can span a maximum of 33,618 functions (20,000 functions on average).  If we consider only functions executed in the buggy scenario, the search space reduces to an average of 234 functions. However, 234 functions still represent a substantial context length for an LLM to analyze effectively. 
Our execution-guided bug localizer further refines this by identifying functions that are uniquely executed in the buggy scenario. This reduces the search space to an average of only 52 functions and a median of just 11 functions. \textbf{This highlights the significant reduction of search space by  \textit{execution-guided bug localizer}.}

\subsection{How similar are \tool's patches to human-written patches?}
\begin{figure}
    \centering
    \includegraphics[width=0.7\linewidth]{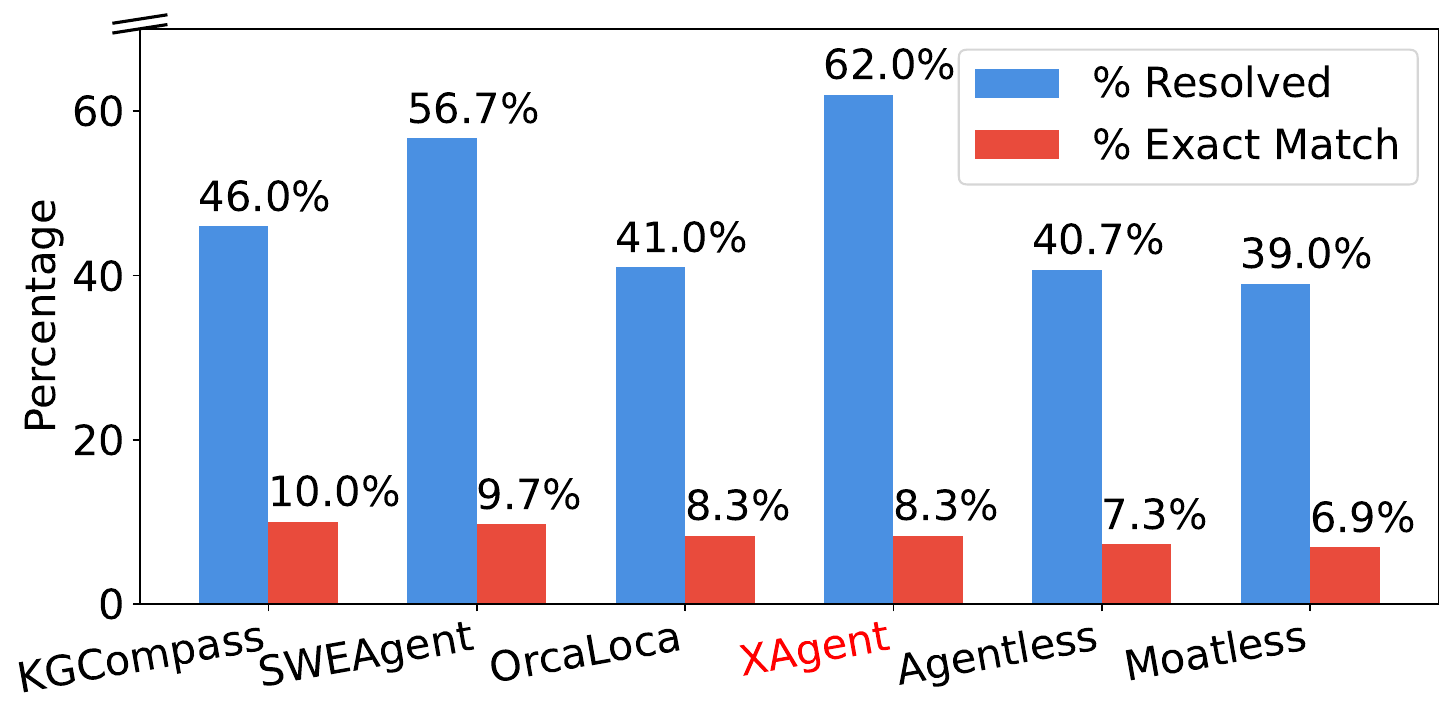}
    \vspace{-5pt}
    \caption{Similarity of generated patches to human-written patches.}
    \label{fig:exact_match}
\end{figure}
As we use Claude Sonnet 4 as our LLM backbone, we investigate whether \tool's performance stems from architectural innovation or potential memorization of ground-truth patches. Similar to prior work~\cite{yang2024unveiling}, we analyze the exact match rate between \tool-generated patches and the golden (human-written) patches.
The exact match rate measures the percentage of instances where the generated patch is identical to the ground truth patch (ignore comments and empty lines).
Fig~\ref{fig:exact_match} presents the \%Resolved and \%Exact match for \tool and five baseline methods \cite{kgcompass,yu2025orcaloca,yang2024sweagent,agentless,antoniades2025swesearch}. 
\tool maintains an exact match rate of only 8.3\%, lower than KGCompass, SWE-Agent, and OrcaLoca.
Upon manual inspection of the exact match cases, we found that most of them involve simple edits of only 1-2 lines of code. These simple cases naturally have fewer solution alternatives, making exact matches more likely regardless of the approach used. For more complex bugs requiring multi-line or multi-function change, \tool consistently generates alternative solutions that differ from human-written patches but achieve functional correctness.
\textbf{These findings indicate that \tool{} does not rely on memorizing ground-truth patches, but instead produces functionally correct solutions, particularly for complex bugs with multiple valid fixes.}

\subsection{How long does \tool take to resolve an issue?}
Execution time is an important factor for understanding the practical behavior of our agentic issue resolution approach.
Our new components (i.e., the executed-guided bug localizer and context-aware validation test) may cost additional computation time.
On average, \tool takes approximately 8.8 minutes to resolve an issue.
The workflow is dominated by two agentic phases: reproduction (181.8s, 34.5\%) and patch generation (240.8s, 45.6\%), which together account for 80.1\% of the total time.
This behavior is expected for agentic issue resolution approaches, as these phases involve iterative LLM interactions with tools and execution environments.

In contrast, our execution-guided localization component (including execution tracing, call tree comparison, and function ranking) collectively consumes only 49 seconds (9.3\% of total time). This demonstrates that our localization approach is highly efficient, adding minimal overhead while providing accuracy improvements. The test augmentation phase takes 56 seconds (10.6\%), which is reasonable given that it generates multiple validation tests to ensure comprehensive patch evaluation.
\textbf{These results indicate that both the execution-guided bug localizer and the context-aware validation test augmentor in \tool are computationally efficient, contributing meaningful performance improvements without significantly increasing overall runtime.}

\section{Limitations}

\pick{SWE-Bench-lite may contain under-specified issues and flaky tests. While SWE-Bench-verified offers manually-verified 500 instances, most baselines use SWE-Bench-lite.
Hence, we focus on SWE-Bench-lite for a fair comparison in our experiment~\cite{verified}.
Nonetheless, these two datasets share an overlapping subset of 93 instances.
Hence, we examine the generated patches of our \tool and TRAE~\cite{traeresearchteam2025traeagent} (i.e., the state-of-the-art agent on the SWE-bench-verified leaderboard). 
We find that \tool and TRAE both achieve the same \%Resolve of 82\% ($\frac{76}{93}$ in this subset.) 
Notably, TRAE is an ensemble method that combines the reasoning of multiple models (Claude Sonnet 4, Opus 4, Sonnet 3.7, and Gemini 2.5 Pro), which may require a higher cost than ours.\footnote{Unfortunately, TRAE did not provide the actual cost of generation. Hence, we cannot compare the cost.}
This analysis provides evidence of comparable performance of our \tool{} between the two datasets.}


The effectiveness of \tool{} relies on the quality of its agent-generated reproduction scripts. While our agent successfully generates buggy and non-buggy scripts pairs for 93\% of evaluated issues. The failure may come from under-specified issues or feature requests. The remaining 7\% must fall back to LLM localization. Furthermore, while techniques like spectrum-based fault localization~\cite{tarantula,ochiai} are more thorough, they often require comprehensive test suites that are rarely available for newly reported bugs \cite{zhang2024autocoderover}. \pick{Therefore, \tool{} can only leverage existing available information in the code to generate reproduction scripts.  
}


\pick{
LLMs may exhibit inherent randomness in their outputs, particularly at higher temperature settings, which can affect reproducibility. Following prior work~\cite{aggarwal2025dars,experepair,kgcompass}, we use low temperature values of 0 and 0.5 for reproduction and patch generation to ensure reproducibility.
Nonetheless, we use a high temperature of 0.8 for the validation augmentor to generate more diverse validation tests. As LLMs may exhibit inherent randomness in this setting, we assess result consistency by repeating the validation generation three times at a temperature of 0.8. Based on cosine similarity over text embeddings generated by OpenAI’s text‑embedding‑3‑large model, the generated validations show an average pairwise similarity of 91.2\% (SD = 0.038) across the runs.
These results indicate that, despite the higher temperature, the model still produces reasonably consistent outputs. To further support reproducibility, we make our complete implementation and experimental setup publicly available at~\cite{xagent}.
}


\section{Conclusion}
\label{section:conclusion}

This work introduces \tool, an execution-guided agentic framework for GitHub issue resolution that advances automated software issue resolution by integrating dynamic program execution with context-aware test generation for patch validation.
By comparing failing and successful executions, \tool~localizes the exact buggy function accurately.
In addition, the context-aware test augmentation enhances validation coverage and reduces overfitting to issue descriptions.
Our empirical evaluation on the SWE-bench-lite dataset shows that \tool~achieves a resolve rate of 62.0\% and a function localization accuracy of 72.8\%, outperforming SOTA approaches (e.g., ExpeRepair, Refact Agent, and SWE-Agent), while maintaining a 37\% lower cost.
These findings demonstrate how moving beyond static descriptions toward execution‑guided reasoning enables LLM‑based agents to deliver more reliable and workflow‑aligned automated software maintenance.

\section{Data Availability}
The source code and experimental details are publicly available at an anonymous GitHub repository~\cite{xagent}.


\section*{Declarations}
\paragraph{Funding} No funding was received to assist with the preparation of this manuscript.
\paragraph{Competing interests} The authors have no competing interests to declare that are relevant to the content of this article.
\paragraph{Ethics approval} Not applicable.


\bibliographystyle{spbasic}      
\bibliography{sample-base,bib}

\end{document}